\documentclass[11pt]{article}
\usepackage{vietnam,epsfig}

\def\be{\begin{equation}}
\def\ee{\end{equation}}
\def\bea{\begin{eqnarray}}
\def\eea{\end{eqnarray}}

\begin{document}
\vspace*{4cm}
\title{Dark Matter on small scales; Telescopes on large scales}

\author{Gerard Gilmore}

\address{Institute of Astronomy, Madingley Rd, Cambridge CB3 0HA, UK}

\maketitle

\abstract{
This article reviews recent progress in observational determination of
the properties of dark matter on small astrophysical scales, and
progress towards the European Extremely Large Telescope.
Current results suggest some surprises: the central DM
density profile is typically cored, not cusped, with scale sizes never
less than a few hundred pc; the central densities
are typically $10-20$GeV/cc; no galaxy is found with a dark mass halo less
massive than $\sim5.10^7M_{\odot}$. We are discovering many more dSphs, which
we are analysing to test the generality of these results. The European
Extremely Large Telescope Design Study is going forward well,
supported by an outstanding scientific case, and founded on detailed
industrial studies of the technological requirements.
}

\section{Dark Matter on small scales}

Are dSph haloes cusped or cored? Debate continues to rage about
whether the cusped haloes always created in CDM simulations are in
conflict with observations of rotation curves of Low Suface Brightness
galaxies. The gas-free nature of dSphs makes them kinematically clean
systems in which to test theoretical predictions.  Stellar velocities
may also be used to place constraints on the steepness of any possible
central cusp, whether due to a black hole, the intrinsic physical
properties of the CDM\cite{T02}, or possibly
even CDM as modified by a central black hole\cite{rg}.

In the last few years several groups have obtained large kinematic
data sets, determining the line-of-sight velocity dispersion across
teh face of many of the local dSph galaxies. These data, together with
surface brightness profiles defining the scale length on which the
light is distributed, allow standard stellar velocity pressure {\sl vs}
gravitational pressure gradient analses through application of the
Collisionless Boltzmann equation, or more commonly the moment
equations known as Jeans' equations.

As a general result, in all cases with sufficient data we rule out
(King model) mass-follows-light models. King models are not an
adequate description of these galaxies, all of which have high
mass-to-light ratios (in solar visual band units), and extended
dark matter halos.  Even in the inner regions mass does not follow
light, while including outer data commonly we find a most likely
global mass to light ratio which is very high, being for Draco $\sim
440$, 200 times greater than that for stars with a normal mass function
(Fig.~1).  The Draco halo models favoured by the data contain
significant amounts of mass at large radii, leading to the observed
flat to rising velocity dispersion profiles at intermediate to large
radii.

Figure~1 summarises Jeans equation models for several of the dSph,
with in each case the simplest possible assumptions (isotropic
radially-constant velocity distribution). It is apparent that the
models are invalid at large radii, where an unphysical oscillation in
the mass profiles is evident. In the inner regions however the fit to
the data is good. In each case, a core-like mass distribution is
preferred. The well-known and irreducible feature of moment analyses
applied to a collisionless sytem, when there is no equation of state
to relate presure and density, however, means it is possible, by adding a
radially-variable stellar anisotropy (essentially an extra stress function)
to the fit, to fit steeper cusp-like central mass distributions.

\begin{figure}
\begin{center}
\psfig{figure=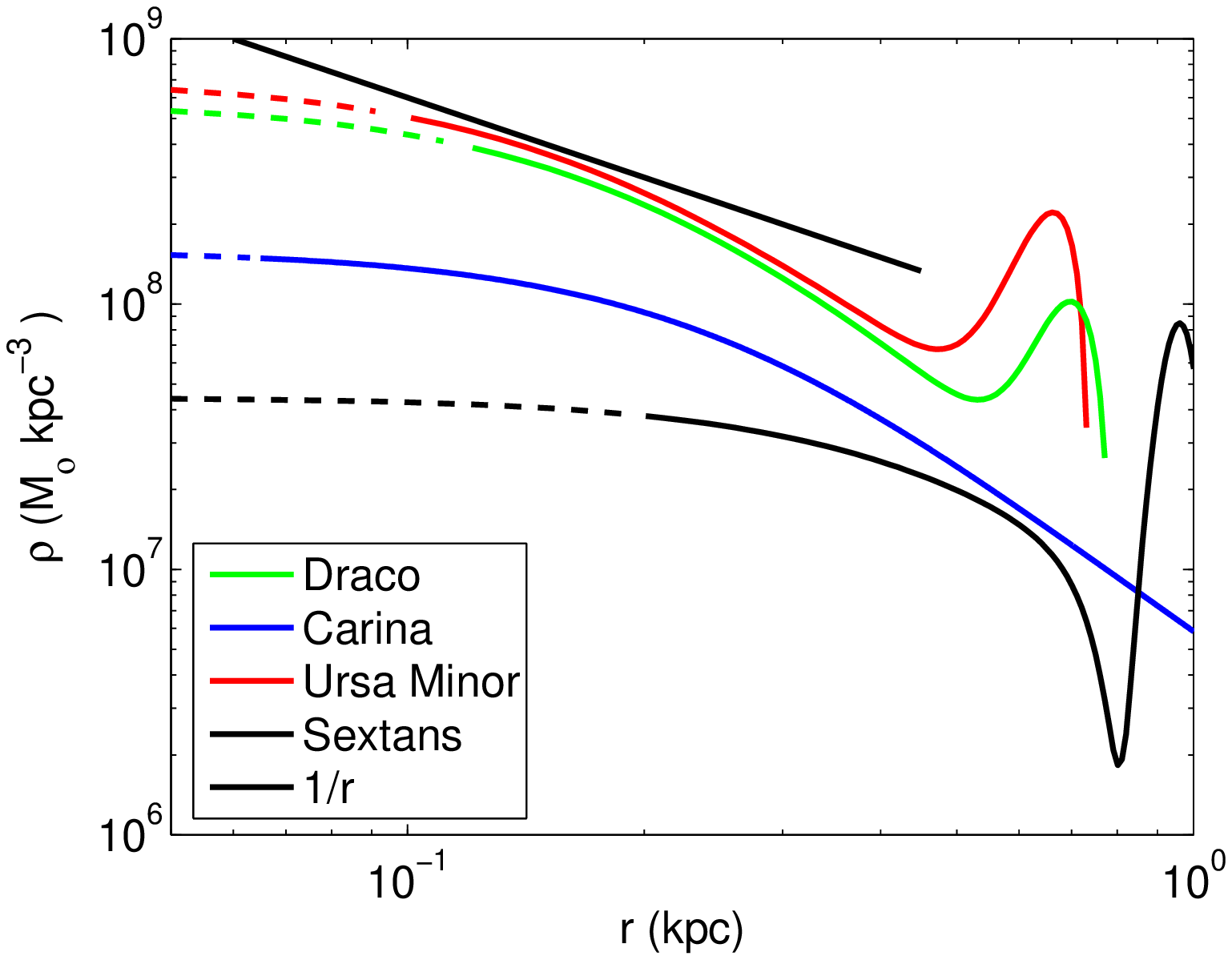,height=2.5in}
\caption{Derived inner mass distributions from Jeans' eqn analyses for
four dSph galaxies. Also shown is a predicted $r^{-1}$ density
profile. The modelling is reliable in each case out to radii of log
(r)kpc$\sim0.5$. The unphysical behaviour at larger radii is explained
in the text. The general similarity of the four inner mass profiles is
striking, in all of shape, length scale and normalisation. This figure
is from Gilmore et al (2006).
\label{fig:DM}}

\vskip 1.5cm
\psfig{figure=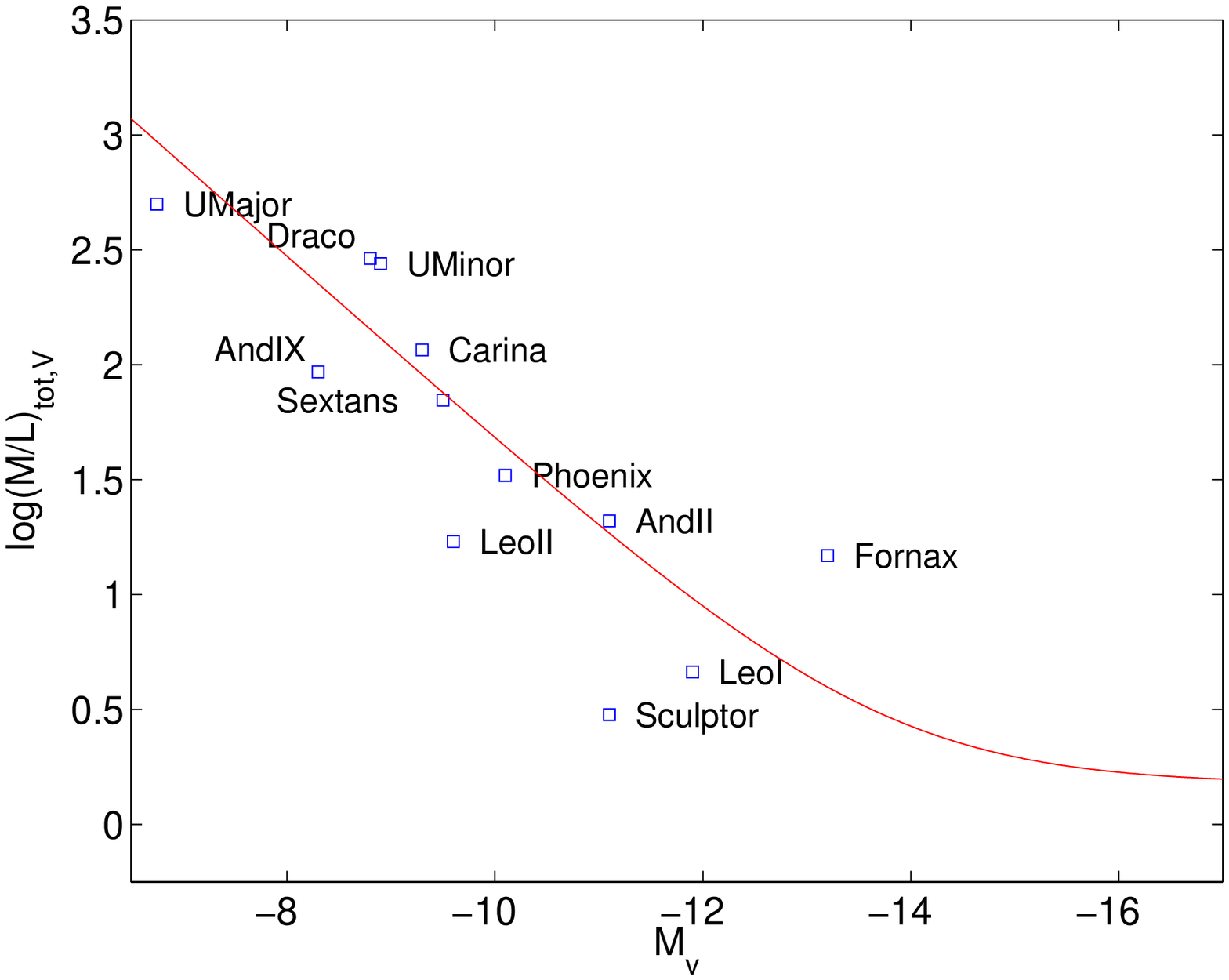,height=3.5in}
\vskip 1.0cm
\caption{The Mateo plot (cf Mateo 1998): 
Mass to light ratios {\it vs} galaxy absolute  V magnitude
for some Local Group dSph galaxies. The solid curve shows the relation
expected if all the dSph galaxies contain about $4\times10^7$ solar
masses of dark matter interior to their optical radii.
This figure is from Gilmore et al (2006).
\label{fig:mateo}}
\end{center}
\end{figure}

The profiles in figure~1, derived by Jeans' equation analyses, and the
 correlation in figure~2, when taken together, 
 illustrate two of our basic results. 
In every case, the simplest analysis favours cored mass
distributions. While cusped mass distribution can usually be fit to
the data, in at least one case, UMi, there is very strong direct
evidence that a cusp model is inadequate to explain all the available
information. The conservative assumption is therefore that all the
mass profiles are indeed cored, and are significantly shallower than
$r^{-1}$.

Secondly, all the dSph we have analysed to date show very similar, and
perhaps surprisingly low, central dark matter mass densities, with a
maximum value of $\sim 5 \times 10^8 M_{\odot} kpc^{-3}$, equivalent
to $\sim 20$GeV/cc. Interestingly, the rank ordering of the central
densities is in inverse proportion to system total luminosity, with
the least luminous galaxies being the most dense. This is of the
opposite sign to some CDM predictions. 

It is apparent from Figure~2 that there is remarkably little spread in
mass within the optical boundary
apparent among the galaxies with absolute magnitute fainter than
$\sim -11.$ This relation was considered until recently to be a minor
curiosity, since it covered the dynamic range only from M$_V \sim -13$
to $-9$, a mere factor of forty or so in luminosity, and included only
8 galaxies. However, the recent analysis\cite{K05} of the
newly-discovered extremely low luminosity dSph galaxy UMa has extended
the validity of the relation by another two magnitudes, now a factor
of $\sim200$ in luminosity, and to total mass-to-light ratios in
excess of 1000. 

The results of figures 1 and 2 are explicable if there is an intrinsic
minimum scale length - about 100pc - an intrinsic maximum central mass
density -- about $10-20$GeV/cm$^3$ - and a similar universal mass
profile. The total mass, simply the product of these, is then
naturally constant. These results are described and discussed further
in Gilmore et al (2007).

Several other new very low luminosity dSph satellite
galaxy candidates have been discovered in the last few months, while
new studies of several known galaxies have recently been
completed\cite{K06}. It will be very interesting to see if these dynamical
studies strengthen or disprove this apparent trend.

\section{The European Extremely Large Telescope}

Several teams worldwide have begun development of the next
generation of Extremely Large ground-based Telescopes (ELTs). A range
of designs and telescope sizes from 25-45m are being considered, with
the larger apertures under study in Europe.

In writing the science case for an ELT, we are in an unusually
advantageous position. The present generation of ground-based
telescopes, complemented by HST and other satellites, have
revolutionised our view of the Universe, illustrating the power of
large telescopes through performance, and have produced a wealth of
fascinating questions that only the vast collecting area and high
spatial resolution of an ELT will be able to answer. These questions
cover areas across planetary science, astronomy and cosmology.  They
range from long-term modeling of weather patterns in Solar System
planets, through direct imaging of Earth-like bodies around other
stars, to understanding the complete formation histories of galaxies,
particularly including the first objects, the role of reionisation,
and the development of dark-matter structures.

\begin{figure}
\begin{center}
\psfig{figure=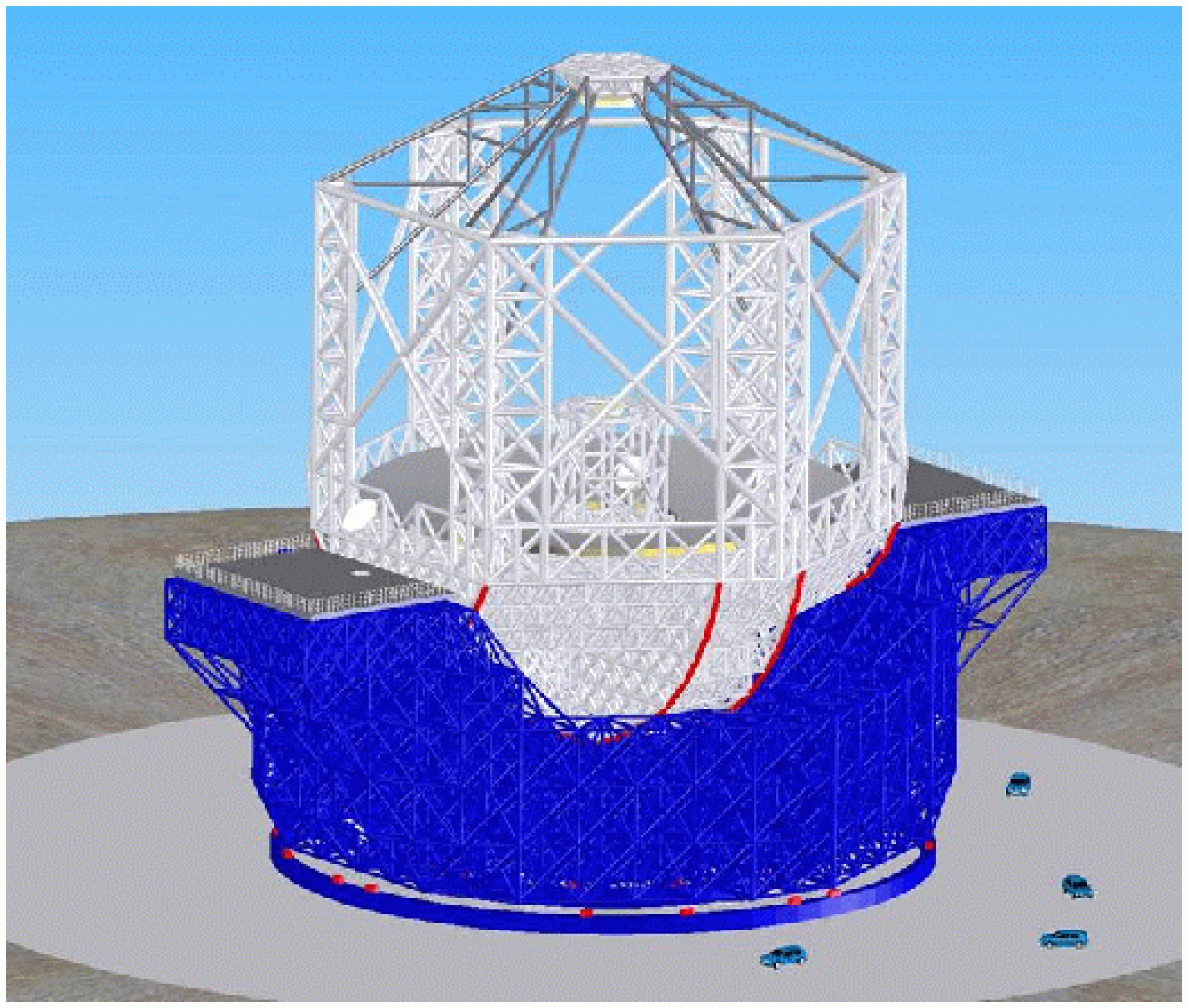,height=4.5in}
\caption{An outline design for a 42-m E-ELT, as of late
  2006. Optimisation and development continue through ESO, national
  agency,  and EC, funded projects.
\label{fig:eelt}}
\end{center}
\end{figure}

Cutting-edge telescopes are immensely flexible tools that can be
turned to many different projects.  This flexibility means that some
of the most exciting discoveries will be those that one cannot 
predict before the instrument is built. Thus, for example, the majority
of the science highlights of the first ten years of the Keck
telescopes' operations -- such as their part in the distant supernova
observations that led to the discovery of dark energy -- were not
featured in the list of science objectives prior to the telescopes'
construction. However, even as written, the science
case for an ELT is spectacular. Not only will it allow us to 
address a wide range of already-posed key scientific questions, but it
will also provide the complementary data that will unlock the full
potential of facilities at other wavelengths, especialy JWST.

The ELT will allow study of planets orbiting other stars -- direct
detection of extra-solar planets and a first search for bio-markers
(e.g water and oxygen) in nearby systems may be feasible with an ELT.
Mapping orbits of gas giants, determining their composition, albedos
and temperatures will be a first step on the way to the more
challenging exo-earth observations. Detailed analysis of the formation of
planetary systems and protoplanetary disks in nearby star-forming
regions will also become possible

Resolved stellar populations studies will extend from studies of
individual stars so far possible only in our Galaxy and its satellites
to a representative section of the Universe, reaching (with the
largest ELTs) the Virgo cluster of galaxies. This will provide information
on how galaxies form, as the ages and compositions of stars reflect past
histories. Massive Black Hole demography will be extended through
dynamical analysis of circum-nuclear regions of galaxies, to establish
whether properties seen in AGN hold also for dwarf galaxies, mapping
Black Hole formation and evolution across a wide mass-range.
Star formation histories across the Universe will be quantified: when
did the stars form? Using the fact that high-mass stars soon die in
supernova explosions, it is possible to deduce the number of stars
that have formed at each redshift, tracing star formation back to
re-ionization.

The dynamics and kinematics of galaxies and their sub-galactic
satellites within large dark matter haloes can be traced with an ELT
out to redshifts of about 5. Thus we can observe the build-up of such
dark-matter structures in the process of formation. Similar supernova
observations to those used to determine the star formation history will
be used to probe on empirical grounds cosmological models for the
nature of dark energy out to the earliest epochs. A first generation
of objects providing the necessary UV photons to re-ionize the
hydrogen in the Universe must have existed. An ELT will distinguish
between candidates: QSOs, primordial stars, SNe.  The brightest
earliest sources (GRBs, SNe, QSOs) are ideal to probe the high
redshift interstellar and intergalactic medium.

In some science cases there is a continuum of improvement as telescope
diameter increases, but there are also a few important critical points
where a telescope above a certain size enables a whole new branch of
study. Three key scientific drivers are used to optimise the science
return as a function of aperture. These are (1) detection and study of
extra-solar planets (2) study of galaxy formation via observations of
resolved stellar populations, quantifying the role of dark matter and
(3) the early universe and the first objects, quantifying the role of
dark energy.

We already know that the up-coming generation of space observatories
(such as Herschel and XEUS) operating at wavelengths from the far
infrared to X-rays, will need the kind of complementary data that only
an ELT can provide. The synergy between ground and space-based
observations in the optical and near-IR has been clearly demonstrated
by projects at the forefront of research which required the
combination of HST and
current 8-10m telescopes in order to make their discoveries: for
example the study of distant supernovae and the discovery and
follow-up of Lyman-break galaxies both make use of HST for imaging,
and large telescopes for spectroscopy.  For many observations a
telescope's ability to detect faint sources scales as $D^2$ and the
time to carry out a given observation scales inversely as $D^4$ (where
$D$ is the primary mirror diameter). Their great aperture means that
ELTs are able to compete with space-based telescopes despite the
reduced background in orbit. For high-resolution spectroscopic
applications, ground-based ELTs can compare in performance with, or
significantly out-perform, the (e.g.) JWST in the transparent
atmospheric windows out to at least 4microns. Even in imaging mode,
ELTs larger than 20m compare in performance to JWST at all wavelengths
shorter than 2.5micron. Already images from HST's Advanced Camera for
Surveys reveal objects that are too faint for the largest existing
telescopes to obtain spectra. The advent of the JWST, scheduled for
launch in 2013, will only serve to increase the opportunity.

The various European projects (EURO-50, OWL) have been brought
together into a single European project, led by and largely funded
through ESO, with significant national and EC support. Figure~3
illustrates the current 42-m aperture default design. Project updates
can be found at www.eso.org/projects.

\end{document}